# Spacecraft at small NEO

D.J. Scheeres
Department of Aerospace Engineering
The University of Michigan
Ann Arbor, MI 48109-2140
1-734-615-3282
scheeres@umich.edu

**Abstract**

The impact risk has been sharply curtailed for the largest km-sized Near Earth Objects (NEOs) through a concerted period of observation spanning the last decade. Thus the risk of impact has now migrated to the smaller and more numerous members of the Near Earth Object (NEO) population. Characterization and mitigation missions to these smaller objects become more complex from a spacecraft operations perspective, even as the prospects for nudging such lower mass bodies off course become more feasible. This paper details the challenges for spacecraft operations at small bodies and indicates areas where more research and analysis of this problem is needed.

## Overview

NASA has devoted significant attention to the scientific investigation of Near Earth Objects over the last decade. This support comes through a variety of programs with a diverse set of motivations – yet all sharing a partial focus on NEO. Some particular examples include observational Astronomy programs with support for the discovery and tracking of NEO, the Planetary Geology and Geophysics Program supporting the study of fundamental geophysical and dynamical evolution of these bodies, the Discovery program with the NEAR mission to asteroid Eros, and the Discovery Data Analysis Program with its support of the analysis of the data returned from that mission. This broad support has allowed for a huge increase in our understanding and characterization of these peculiar solar system objects, to the point where it now becomes feasible to seriously consider how such bodies could be controlled or deflected should one be found on a high probability impact trajectory with the Earth.

In contrast to these programs dedicated to the scientific understanding and cataloging of these objects, there have been no dedicated programs to the study of controlled motion on or about these bodies. Historically this made sense, as the specific problem of spacecraft motion about these bodies was tied directly to specific missions with their particular scientific goals. However, when considering the development and characterization of space-based technologies for possible mitigation or evaluation missions to an open set of such small bodies it becomes apparent that a significant gap exists in a universal understanding of the issues that such missions will face in the small body dynamical environment. This is especially true for the class of missions that can be seriously considered for mitigation missions: small asteroids with significant levels of interaction between the orbiting craft and the small body itself.

This white paper will present the issues that currently exist for such a class of missions and propose that a modest program of research and technological development be started to address these issues. An initial program on the order of $500K/year for 3-5 years would enable fundamental issues to be researched, navigation techniques to be developed, and the creation of a community of expertise that can be used to rapidly develop and evaluate such missions should the need ever arise. Unlike specific technologies that may be proposed to perform a mitigation or evaluation mission, the theory, understanding and applications of dynamical motion on and about small bodies

falls into the category of a core capability that supports all proposed missions to these bodies.

**History**

To date there have only been two space missions that directly entered into the asteroid environment, NASA's Near Earth Asteroid Rendezvous (NEAR) mission to asteroid Eros and the JAXA (Japanese) mission Hayabusa to asteroid Itokawa. The mission designs used by each mission, and the asteroids visited by each mission, are fundamentally different. NEAR was centered around an orbital mission design whose largest concern was to avoid destabilizing interactions with the gravity field of the asteroid [1]. Hayabusa, on the other hand, never went into a full orbit about its target asteroid and instead relied on active thruster firings to null out the attraction of the asteroid [2]. In fact, the mission design approaches used for NEAR could not have been used (easily) for Hayabusa, and the Hayabusa hovering approach could not have achieved the same high level of science return at Eros [3]. The difference between these approaches is fundamentally controlled by the size of the asteroid visited, with Itokawa two orders of magnitude smaller in extent than Eros.

Thus, it is tempting to adopt the Hayabusa hovering concept for missions to similarly small (< 300 meter across) NEO. A more careful consideration reveals some significant problems with the approach, however. Specifically, one of the prime needs from a characterization or mitigation mission will be to precisely determine the mass distribution and orbit of the body to support subsequent mission design or to precisely determine the body's orbit and any deflection in its trajectory. The Hayabusa mission approach, however, did not allow for such high precision determinations to be made, and was only able to determine the total mass to about 5% and was not able to draw any conclusions on the homogeneity or in-homogeneity of the mass distribution, despite descending to the surface twice [4,5]. Additionally, the orbit determination of the spacecraft relative to the asteroid was not able to provide any improvement to the asteroid ephemeris beyond what the ground-based radar observations had determined previously. These limitations are tied directly to the hovering approach used by Hayabusa, as the frequent firing of thrusters destroys crucial information about the motion of the spacecraft in that environment from which mass distribution and ephemeris improvements can be inferred. This is in contrast to the NEAR mission with its orbital mission design approach from which extremely precise measurements of the gravity field and rotation state of the body were derived, and for which the asteroid orbit was determined to a sub-meter level of precision [6,7].

In terms of the scale of the spacecraft and asteroid, neither of these missions approach the potential size issues that may exist between small asteroids and the large spacecraft that may be sent to them for mitigation operations. Additionally, only minimal information has been gleaned concerning the interaction of a spacecraft with an asteroid surface, with the clearest data coming from the Hayabusa mission [8], although the results of these measurements is inconclusive.

**Challenges**

There are a few fundamental questions that must be addressed in order to properly understand what mitigation approaches can be used to deflect a small NEO.
- Is it possible to carry out NEAR-level precision trajectory navigation at a few hundred meter sized or smaller asteroid?

D.J. Scheeres, The University of Michigan, scheeres@umich.edu

- Is it possible to carry out surface operations on a small body, where every mechanical interaction may raise clouds of pebbles in long-lived orbits about the body?
- Is it possible to guide an impactor traveling at orbital speeds to reliably strike a body a few hundred meters across or less?
- Is it possible to determine the change in orbit due to such an impact?
- What issues will arise when maneuvering a large space structure in close proximity to a small asteroid?

Among the many challenges and open questions that exist, we will briefly discuss three: design of characterization missions to a small NEO, determining the change in the orbit of an NEO due to a mitigation event, and challenges of operating a large spacecraft on and about a small NEO. Clearly, the set of problems and issues that need to be addressed are larger than these.

**Characterization Mission Challenges**

Characterization missions are a pre-requisite for any impact mitigation operation, as it is necessary to determine the basic physical parameters of the body, measure its mass distribution by determining its gravity field and potentially its inertia moments, carry out experiments to determine its surface composition and mechanical properties, among other items in order to prepare an appropriate mitigation technique. Due to their small size, however, orbital operations about such bodies may be difficult, depending on the spacecraft properties and the NEO density and size. Even if an orbiting mission can be carried out, the large perturbations from the solar radiation pressure (SRP) and asteroid non-spherical gravity field can severely constrain and destabilize an orbit. As evident from data returns from the Hayabusa mission, a hovering approach is not sufficient to determine any higher degree mass distribution.

The difficulty of carrying out a precision scientific mission about a small asteroid are related to the force perturbations active in that environment. These perturbations consist of the solar gravity and tidal perturbation, solar radiation pressure perturbation, and the asteroid non-spherical gravity field perturbations. Each of these effects can destabilize a small body orbiter, either by itself or by acting in conjunction with each other. For the size of asteroids and orbiters being considered the solar gravity and tide perturbation is in general negligible and need not be considered in the analysis. The effect of the Solar Radiation Pressure (SRP) and the asteroid gravity field are both directly important and place strict constraints on the orbits that can be feasibly flown by the spacecraft.

The SRP perturbation places several constraints on the orbit design, including the orbit semi-major axis, eccentricity, inclination, argument of periapsis and longitude of the ascending node. First, it places an upper limit on the semi-major axis of the asteroid orbiter, as orbits that are too large can be directly stripped out of orbit about the asteroid and escape into orbit about the sun. The upper limit for such orbits can be approximately bounded using basic analysis and dynamical systems techniques [9]. The specific boundary between a bound orbit and an unbound orbit is not a sharp dividing line, and for a specific asteroid, spacecraft, and orbit orientation must be determined in a practical sense. Still, approximate bounds can be found that are conservative in general, meaning that an orbit chosen according to them will be bound.

D.J. Scheeres, The University of Michigan, scheeres@umich.edu

In addition to orbit semi-major axis, the SRP perturbation only allows for a narrow class of orbits that can be safely flown without needing frequent correction maneuvers [10,11]. These orbits have a characteristic orientation, with the orbit plane lying in the terminator plane and the orbit angular momentum either pointing towards or away from the sun. Then, due to the SRP forces acting on the spacecraft the orbit plane is forced to precess at the same rate as the asteroid moves about the sun, even accounting for an elliptic helio-centric orbit. Thus, these orbits will naturally follow the sun as the asteroid orbits about the sun. In this sense they are sun-synchronous orbits, although they rely on a completely different effect than Earth sun-synchronous orbits rely on and they are forced to always lie in the sun terminator plane.

While the SRP perturbation constrains the orbits the spacecraft can be in, the effect of the asteroid gravity field will in general place additional limits on these orbits. As a perturbation there are two main effects that arise from the mass distribution, precession of the orbit plane due to the asteroid oblateness and resonant interactions with the rotating gravity field due to its equatorial ellipticity. Both of these interactions vary inversely with the distance to the central body, and both of them can destabilize an orbit. The oblateness can cause the orbit plane to precess out of the terminator plane and thus cause the SRP to excite oscillations in eccentricity. The ellipticity can cause fluctuations in the orbit semi-major axis and eccentricity, inducing a chaotic variation in these quantities that can lead to impact or escape. Both of these perturbations are functions of the asteroid rotation pole relative to its heliocentric orbit, rotation period, and shape. Since the rotation pole and shape are often not available prior to rendezvous it becomes difficult to develop globally valid limits for orbit stability that do not directly depend on these parameters. A conservative design philosophy leads to overall lower limits to the minimum orbit radius about the body. For small NEO, these minimum orbit radii can be approximately equal to the maximum semi-major axis for capture relative to the SRP perturbation. When confronted with these situations it becomes necessary to better understand and characterize the orbital dynamics of spacecraft about such small bodies.

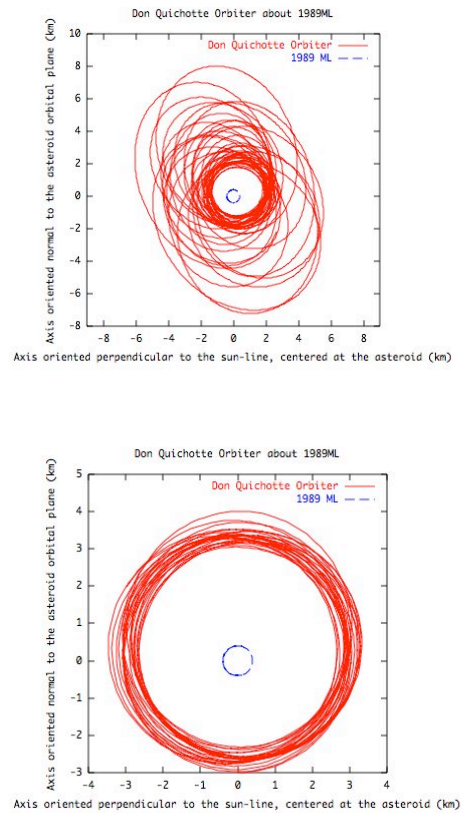

Figure 1: Example orbits about 1989ML. Left, initial orbit radius of 2 km, at a distance where resonance effects from the mass distribution are important. Right, initial orbit radius of 3 km, outside the reach of the mass distribution effects.

Current theory does not adequately explain the effect of joint perturbations on the orbit

D.J. Scheeres, The University of Michigan, scheeres@umich.edu

dynamics and hence there exist open issues for the reliable design of missions to such small bodies. As shown in Fig. 1, the effect of mass distribution can destabilize motion and lead to impact or escape of the orbiting spacecraft. On the other hand, the presence of a minimum radius for a safe orbit directly conflicts with the goal of measuring the mass distribution. This sets up a design optimization problem in the developing a mission plan that simultaneously guarantees orbital safety and appropriate precision in the characterization of the body. Characterizations of such limits, which will be crucial in determining what sorts of missions are feasible for these bodies, can be made but haven't been considered to date in the research community.

For very small NEO, it may not be possible to enter an orbit about the body. For these situations new approaches to the characterization of these bodies must be developed. Hayabusa-style hovering will most likely not be an option, due to the concerns discussed earlier. Instead, it may be possible to perform repeated low-altitude flyovers of the NEO at a variety of solar latitudes in order to measure the net gravitational attraction from the body at different geometries. Such an approach has not been considered or proposed previously, but has heritage in the Galileo and Cassini flybys of planetary satellites and their use to determine low-order gravity coefficients for these bodies. To be useful for NEO characterization purposes an extension of these approaches must be developed.

**Measuring the deflection of an NEO**

At the core of any impact mitigation mission is an ability to, first, precisely determine the NEO orbit and then, if necessary, measure the change in the NEO orbit due to the mitigation event. Extrapolating from the NEAR mission, it becomes clear that once a body is in a precise orbit about an NEO, and given sufficiently accurate navigation sensors and a "clean", maneuver free environment, it is possible to perform extremely precise orbit determination on the NEO. In principle this orbit determination works by precisely determining the spacecraft trajectory about the NEO and also precisely determining the spacecraft trajectory in the solar system. Since the solar system trajectory of the spacecraft is forced to follow the motion of the NEO, and the NEO is resistant to non-gravitational forces such as solar radiation pressure, it becomes possible to perform extremely precise orbit determination on the NEO [7].

Thus, if a spacecraft can enter a stable orbit about an NEO and can perform sub-meter orbit determination relative to the NEO, it can most likely meet the stringent constraints on measuring the deflection imparted to a small body. The more difficult situations involve when orbiting is not an option, and when no direct characterization mission is sent to the asteroid. It can be argued that the latter case will never occur, in that even if the mass distribution properties of the body are not needed, at the least a precise determination of the NEO orbit will be necessary. The question in this situation is then reduced to what is the minimum capability spacecraft that can be designed to reliably determine the trajectory of an NEO before and after a mitigation event. Due to the strong constraints that exist on mission design due to the NEO environment, the spacecraft design must be made while incorporating these orbital constraints. This serves as another example of a highly integrated design optimization process in this field that must be resolved.

Further, if plans are made to carry out an impact mitigation mission with a kinetic impactor, similar to the Deep Impact mission, a severe challenge is to reliably and

D.J. Scheeres, The University of Michigan, scheeres@umich.edu

precisely impact the small body. Closed loop navigation approaches used by Deep Impact may not be precise enough, given the much smaller target object. Instead, it may be of interest to use the in-situ NEO orbit determination craft at the small body to serve as a targeting beacon for the approaching impactor.

If an orbiter mission is not feasible, then the challenge is to design a navigation approach that can effectively remove the thruster firings that will be necessary to keep the spacecraft in the vicinity of the NEO. At the least requirements must be placed on the precision of the accelerometers to be used for the spacecraft. In addition, to reduce the costs and dependence on ground navigation an autonomous station keeping and orbit determination capability must be developed.

**Orbit and attitude coupling**

One class of mitigation mission places the spacecraft in close proximity to the NEO, and either utilizes a mechanical attachment to transmit force [12] or utilizes the spacecraft mass itself to tug the body along its path [13]. Spacecraft designed for these operations will, by definition, be large structures in close proximity to the small body. There will be non-negligible coupling between the translational motion of the spacecraft and its rotational motion, especially when precise positioning between the bodies is needed.

Although the problem resembles the rendezvous and docking problem faced by craft such as the shuttle and space station, there are fundamental differences between those problems and the NEO problem. The driving distinction for the NEO problem is the non-negligible mass of the small body. Unlike the docking problem in Earth orbit, the mass of the NEO cannot be ignored and will, in fact, be the largest perturbation acting on the orbit and attitude dynamics. The implications of this are that standard algorithms and classical results cannot be transported to this environment, and the theory of rendezvous and docking must be revisited from the ground up.

As a simple example, consider the stability of gravity gradient spacecraft, a classical problem that has been understood for Earth orbiters for decades. When considering similar orbits for a large spacecraft about a small NEO the stability of the classical gravity gradient solution can be lost, even for a spherical NEO. Basically, the size scale between the two bodies becomes similar enough so that assumptions that can be implicitly made for Earth orbiting spacecraft are violated at a NEO, and situations can arise where there are no stable gravity gradient orientations possible between the bodies. The implications of this result alone are large, as it is common to tacitly assume the stability of such orientations based on more familiar results from our own engineering experience. Thus, just this particular dynamical instability may complicate the implementation of a number of close proximity mitigation approaches and reduce their practicality and feasibility.

For this class of problems there is virtually no research beyond some recent forays into understanding the stability of such situations for natural systems [14]. Thus there is need for fundamental research in this area to expose the basic dynamics of such interactions and to develop a theory for the control and navigation of spacecraft in these situations.

**Conclusions**

There are fundamental unknowns concerning a number of crucial aspects of spacecraft navigation, dynamics and control about small NEO. While there has been some progress in the general theory of such situations, there are many open questions that must be resolved and properly understood in order to critically evaluate different mitigation concepts and subsequently implement them. A modest program over a 3-5 year period could jumpstart research on these issues and, with high probability, resolve many of the most fundamental issues related to such mitigation missions.

**References**

[1] Scheeres, D.J. "Analysis of Orbital Motion Around 433 Eros," The Journal of the Astronautical Sciences 43:427–52, 1995.

[2] Takashi Kominato, Masatoshi Matsuoka, Masashi Uo, Tatsuaki Hashimoto and Jun'ichiro Kawaguchi, "Optical Hybrid Navigation In Hayabusa - Approach, Station Keeping & Hovering," paper presented at the AAS/AIAA Space Flight Mechanics Meeting, Tampa, Florida, January 2006. Paper AAS 06-210.

[3] D.J. Scheeres. "Close Proximity Operations at Small Bodies: Orbiting, Hovering, and Hopping," in Mitigation of Hazardous Comets and Asteroids, (M. Belton, T.H. Morgan, N. Samarasinha, D.K. Yeomans eds.), Cambridge University Press, 2004.

[4] A. Fujiwara, J. Kawaguchi, D. K. Yeomans, M. Abe, T. Mukai, T. Okada, J. Saito, H. Yano, M. Yoshikawa, D. J. Scheeres, O. Barnouin-Jha, A. F. Cheng, H. Demura, R. W. Gaskell, N. Hirata, H. Ikeda, T. Kominato, H. Miyamoto, A. M. Nakamura, R. Nakamura, S. Sasaki, and K. Uesugi. "The Rubble-Pile Asteroid Itokawa as Observed by Hayabusa," Science 312: 1330-1334, 2006.

[5] S. Abe, T. Mukai, N. Hirata, O. S. Barnouin-Jha, A. F. Cheng, H. Demura, R. W. Gaskell, T. Hashimoto, K. Hiraoka, T. Honda, T. Kubota, M. Matsuoka, T. Mizuno, R. Nakamura, D. J. Scheeres, M. Yoshikawa. "Mass and Local Topography measurements of Itokawa by Hayabusa," Science 312: 1344-1347, 2006.

[6] Miller et al., Icarus 155, 3–17, 2002

[7] Konopliv et al., Icarus 160, 289–299, 2002

[8] H. Yano, T. Kubota, H. Miyamoto, T. Okada, D. J. Scheeres, Y. Takagi, K. Yoshida, M. Abe, S. Abe, O. Barnouin-Jha, A. Fujiwara, S. Hasegawa, T. Hashimoto, M. Ishiguro, M. Kato, J. Kawaguchi, T. Mukai, J. Saito, S. Sasaki, and M. Yoshikawa. "Touch-down of the Hayabusa spacecraft at the Muses Sea on Itokawa," Science 312: 1350-1353, 2006.

[9] Scheeres, D.J. and Marzari, F. Spacecraft Dynamics in the Vicinity of a Comet. Journal of the Astronautical Sciences 50(1): 35-52, 2002.

[10] Scheeres, D.J. Satellite Dynamics about Small Bodies: Averaged Solar Radiation Pressure Effects. Journal of the Astronautical Sciences 47(1): 25-46, 1999.

[11] Morrow, E., Scheeres, D.J., and Lubin, D. Solar Sail Orbit Operations at Asteroids. Journal of Spacecraft and Rockets 38(2): 279—286, 2001.

[12] D.J. Scheeres and R.L. Schweickart. "The Mechanics of Moving Asteroids," paper presented at the AIAA Planetary Defense Conference, Orange County, California, February 2004. Paper AIAA-2004-1446.

D.J. Scheeres, The University of Michigan, scheeres@umich.edu

[13] E.T. Lu and S.G. Love, "Gravitational tractor for towing asteroids," Nature 438, 177-178, 2005.

[14] D.J. Scheeres. "Relative Equilibria for General Gravity Fields in the Sphere Restricted Full 2-Body Problem," Celestial Mechanics and Dynamical Astronomy 94(3): 317-349, 2006.

D.J. Scheeres, The University of Michigan, scheeres@umich.edu